# Cosmic meteorology

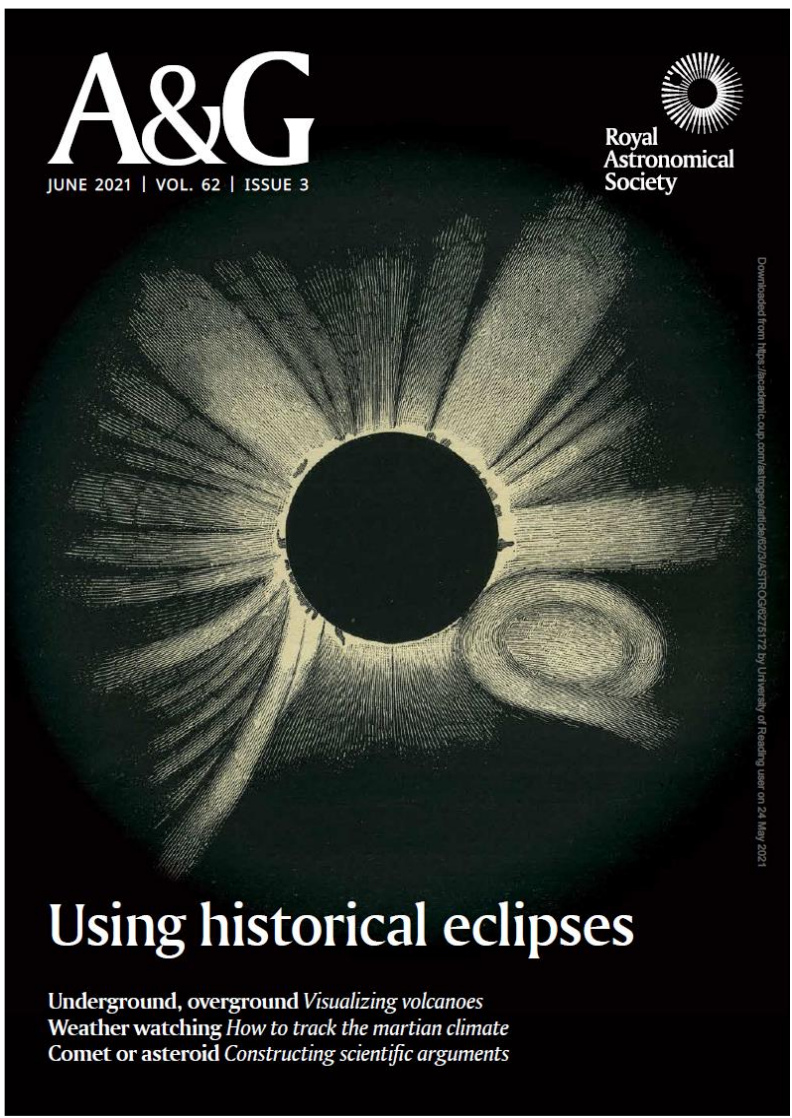

**Cover Image:** An engraving by William Henry Wesley of the total solar eclipse of 18 July 1860, published in Memoirs of the RAS in 1879. This engraving was based on observations by Ernst Wilhelm Leberecht Tempel; others recording the same event also saw the unusual curved structure. See page 3.12 to find out more about how such historical observations have a place in today's space weather research. (RAS)

Mike Lockwood and Mathew Owens discuss how eclipse observations are aiding the development of a climatology of near-Earth space.

The term "cosmical meteorology" was first used by Italian astronomer Giovanni Battista Donati, inspired by the great aurora of 4 February1872.  A century and a half later, we now use his concept – calling it "space weather".  The analogy is a good one because, as in meteorology, an important aim is to build a long-term climatology of our near-Earth space environment to help design and operate facilities that are at risk from the hazards posed.  To do this, space science is adapting many of the techniques that have been developed by meteorologists and climate scientists such as: ensemble forecasts using numerical models, skill scores, cost-benefit analysis, data assimilation and re-analysis.  In the case of space weather, re-analysis means reconstructing past conditions in near-Earth space using a data-assimilation algorithm of some kind to combine a variety of historical observations with predictions made by a model that is driven by historical estimates of the state of the Sun.  The data used to describe the Sun have almost always been sunspot numbers, but we here discuss how eclipse observations of the solar corona could also be of use in constraining the model and so adding to the analysis.

Recent years have produced great advances in our understanding of the long-term variability in the climate of near-Earth space (see box "Space Climate" and Figure 2 at the end of this paper).  We now know that the full range of variability of average conditions over the past 9000 years was also present during the four centuries since telescopic measurements of sunspots were first made by Thomas Harriot in 1610.   In those 400 years, the Sun has varied between the Maunder minimum (circa 1650-1710, when very few sunspots were observed and the few that did appear were, unusually, all in the southern solar hemisphere) (*Eddy*, 1976; *Usoskin et al.*, 2015) and what we now recognise to have been the peak of a grand solar maximum around 1985 (*Lockwood et al.*, 2009).

The reason we wish to build a climatology of near-Earth space is that variations generated by solar activity have the potential to disrupt, damage and degrade modern technology such as broadcast and communication satellites, navigation and radar systems, aircraft avionics and power distribution networks (*Cannon et al.*, 2013).  In addition, both the energetic particles generated by disturbed space weather and galactic cosmic rays (which, conversely, we are shielded from by disturbed space weather) are health hazards for astronauts and humans flying in high-altitude aircraft (*Lockwood and Hapgood*, 2007). In order to build robust, yet cost-effective, systems and operate them safely and reliably, we need a climatology that can aid their design and use. We have space measurements that extend back to the early 1960s and these can be used to build an empirical climatology. The problem is that we now realise that this is interval is not representative of all, or even average, space weather conditions and if we are to plan for the future we need to build a space climatology that does not have that limitation. That means that we have to gain an understanding of what space

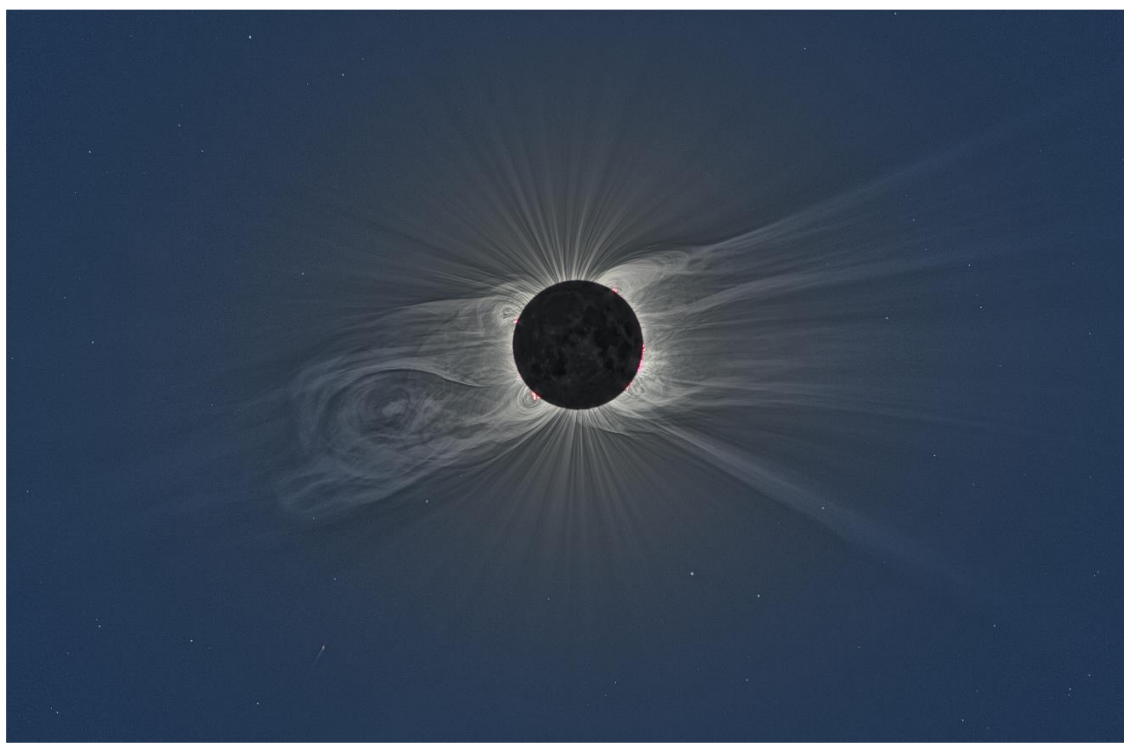

Figure 1. *The fabulously detailed image of the 14 December 2020 solar eclipse captured by Andreas Möller, at Fortin Nogueira, Neuquen, in Argentina and after the outstanding image processing by Miloslav Druckmüller. The corona is a classic sunspot minimum form, but with a clear Coronal Mass Ejection (CME) leaving the Sun down the streamer belt to the left of the image.*

weather was like before the space age began.

From the data that we do have, allied to modelling based on more recent observations by spacecraft, we have been able build up a picture of how space weather has varied over the past 400 years (*Owens et al.,* 2017, *Lockwood et al.,* 2017). However, there is one set of regular observations that are relevant to space weather and could, potentially, help constrain our understanding but has proved hard to exploit. This is the series of reports, sketches and photographs of the solar corona during total eclipses of the Sun. However much has changed in humankind's ability and willingness to observe eclipses in the past 400 years. A key factor in the past was that many eclipses can only be seen from remote locations. Nowadays, eclipse tourism is common and many individuals travel to such remote places to make observations, equipped with photographic and optical equipment and image processing techniques that early observers could not have even imagined. But eclipse observation expeditions could only begin when reliable predictions of when and where eclipses would occur became available.

# Predicting eclipses

Edmund Halley famously made predictions for the 1715 eclipse in the form of a terrestrial map giving the path and times of totality. The timings turned out to be correct to within about 4 minutes, but the observed path to be displaced by about 30 km kilometres from the prediction. Nevertheless, the degree of accuracy was unprecedented and was largely obtained from the application of Newtonian orbital mechanics. Indeed one man who helped Newton formulate the mathematics needed for that theory, Roger Coates, made use of Halley's predictions to observe the eclipse from Cambridge while Halley himself observed it from the old Royal Society premises just off Fleet Street. Halley initiated a very early (maybe the first) "citizen science" project by publishing the predictions in a 'broadside' pamphlet that also encouraged "curious" readers to "observe it … with all the care they can" and to report back to him so that "we may be enabled to predict the like appearances for the future, to a greater degree of certainty." We are not sure how many copies were sold, but an average broadside print run was about 2000. We do know that it sold for 6 (old) pence (approximately £5 in modern prices) and that Halley was sufficiently pleased with the response that he corrected the 1715 predictions and repeated the exercise for the 1724 eclipse.

Several ancient civilizations had pursued eclipse observation and prediction long before Halley. The earliest known records are from China, and many cultures developed methods to predict eclipses from the patterns of occurrence, though they did not understand why they worked and could not predict where they would occur. The earliest discovered historical record giving the length of a 'Saros cycle' is by Chaldean astronomers (a country in the marshlands of south-eastern Mesopotamia that existed between the 9th and 6th centuries BC, after which it, and its people, were assimilated into Babylonia). We now know that this is the interval between the Earth, the Moon and the Sun falling in line. On its own, this knowledge gave a way to predict when eclipses would occur but not where.  The Saros period was known to philosophers such as Aristotle and Hipparchus in Greece, Pliny in Rome and Ptolemy in the part of the Roman Empire that is modern-day Egypt.  It was in ancient Greece that it was established that solar eclipses are caused by the Earth moving into the shadow of the moon. The first accurate prediction of a solar eclipse using this understanding was, at least according to the historian Herodotus, by Thales of Miletus – and if this report is true it is likely that the solar eclipse in question was on 28 May 585 BC. Note that for many centuries predictions were based on Ptolemy's Earth-centred universe.

The renaissance meant that the writings and thoughts of ancient Greek and Roman philosophers were revived and with major increases in the accuracy of astronomical observation and mapping of the Earth it became possible to predict the geographical locations at which totality would be seen as well as when an eclipse would occur. In 1705 two maps were published in advance of the eclipse of 1706 and many astronomers such as de Plantade and de Clapiés in Montpellier and Wurzelbauer and Eimmart in Nürnberg were ready to observe it at the predicted time and place.

# Eclipse expeditions

It has been suggested that the first official eclipse expedition was a group from Harvard who travelled to Maine in October 1780.  If so, this took them behind the British lines during the American war of independence but, sadly, their bravery was not rewarded because an error in their longitude calculation meant that they failed to observe totality. Later, steam ships opened new possibilities for tourism and increasingly enabled eclipse expeditions by interested organisations such as national observatories, astronomical societies and universities.

At the same time, the technology of the observing instruments developed, including the advent of photography and spectroscopy.  The knowledgeable and literate individuals who staffed the expeditions generated detailed descriptions, mercifully less full of analogies and hyperbole than those by the more random selection of observers before predictions made expeditions viable.  (However, long tracts of their reports are often like a tourist's holiday diary!).  Modern-day expeditions are a mixture of enthusiasts, hi-tech observers and image processers, and tourists. The use of polarised light cameras enables images of the corona to be made further away from the Sun than is otherwise possible (a point first noted by *Ranyard*, 1871). These cameras can detect the "K-corona", light that is Thompson-scattered by electrons and ordered by the coronal magnetic field. Without them the K corona would be lost beyond about two solar radii, being of lower intensity there than the "F-corona", light that has been scattered by dust. In addition, advanced image-processing techniques, in particular allowing for the rapid decrease in K-coronal intensity with distance from the Sun, mean that extraordinarily detailed views of the coronal magnetic structure can now be obtained.

From the expeditions, it was realised in the late 19$^{th}$ century that the nature of the solar corona varied over the sunspot cycle. A decadal-scale cycle in sunspots had first been suggested by Christian Horrebow in 1775, based on his observations over the previous 14 years and firm evidence was published by Samuel Heinrich Schwabe in 1843. The infrequent nature of observable eclipses meant that the realization that the form of the corona varied over the sunspot cycle was a slow one. First suggestions appear to be by Pierre Jules César Janssen in 1878, by *Ranyard* (1879) in his survey of coronal forms published in Memoires of the Royal Astronomical Society and by *Hansky* (1897). *Bigelow* (1890) was the first to recognise the solar magnetic field was a key component in the coronal structure. Certainly, the evidence was available by then. For example, the characteristic sunspot minimum form (with a bright streamer belt separating the large dark polar coronal holes) and the sunspot maximum form (with streamers at all heliographic latitudes) was clearly evident in the eclipse sketches, made in 1878 and 1871, respectively by William Harkness, a Scotsman who became an Admiral in the US navy, astronomical director of the Naval Observatory and later president of the American Association for the Advancement of Science. These are shown in parts a and e of Figure 3. It is quite likely that the eclipse of 1878 in the USA was highly influential in this revelation, with Étienne Léopold Trouvelot's famous painting (like that of the same event by Harkness, made at Creston, Wyoming from photographs) also showing a clear and narrow steamer belt. The average $aa_H$ index at the time was as low as it was during the low solar minimum of 2019 and that is indicative of a low open solar flux (OSF), the total magnetic flux that leaves the solar corona and enters the heliosphere (Lockwood et al. 1999, 2014). The modelling by Lockwood & Owens (2014) and Owens et al. (2017) predicts that this OSF at sunspot minimum gives a similar, well-defined streamer belt, as was detected in both 1878 and 2019 (Figures 3e and 3d). Figure 3 also shows the corona at the start, maximum and end of two very similar sunspot cycles, numbers 14 and 24. These images clearly reflect the changes in the fundamental nature of the corona over the sunspot cycles.

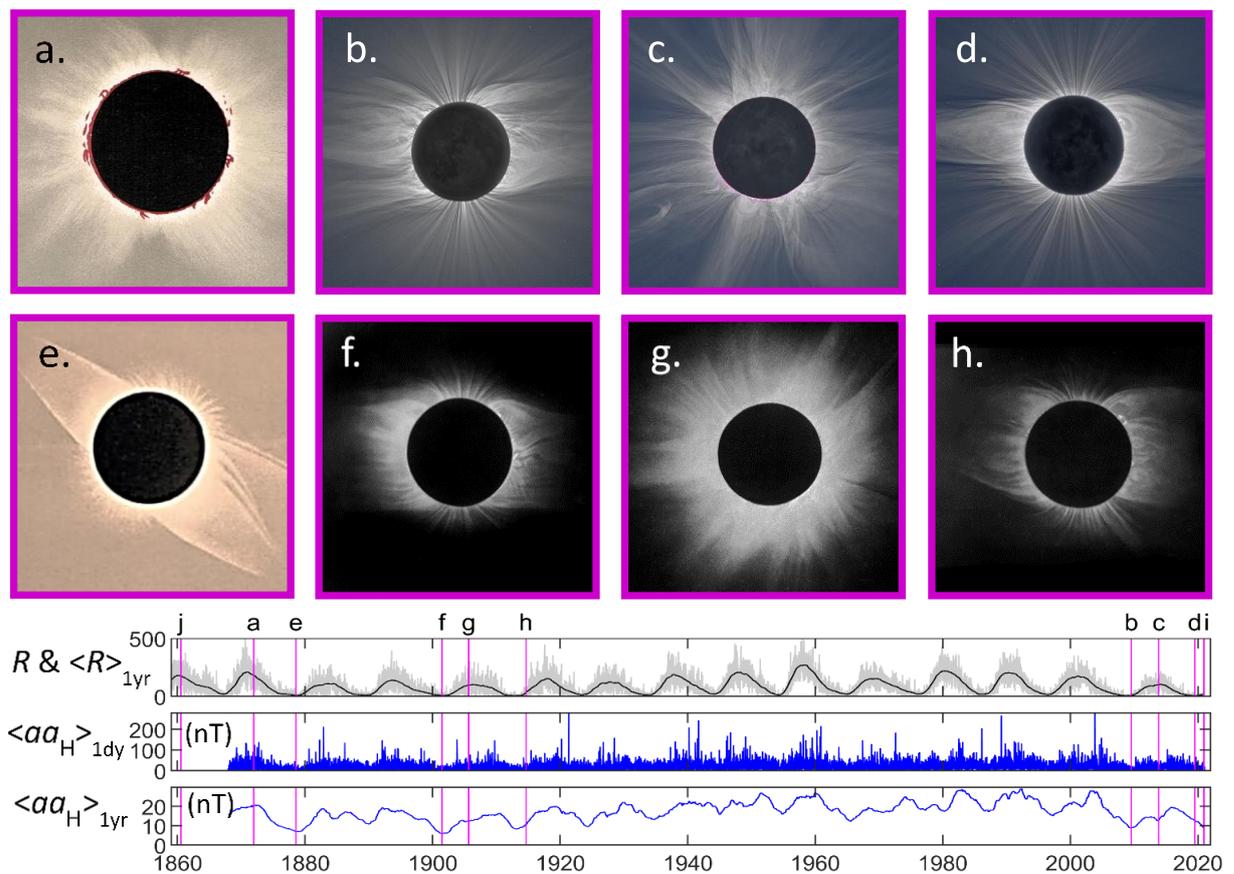

Figure 3. *Eclipse images showing the effect of the sunspot cycle on the corona. Images a and e are paintings of eclipses observed by Admiral William Harkness, the first on 12 December 1871 during an expedition to the South China Sea at sunspot maximum, the second from Creston, Wyoming, USA on 29 July 1878 at the subsequent sunspot minimum. Parts f, g and h are taken at the start, maximum and end of solar cycle 14 and parts b, c and d are taken at the start, maximum and end of a very similar solar cycle 100 year later, cycle 24. The times i and j are for the eclipses presented in Figures 1 and 5. The bottom 3 panels show the time series of the revised international sunspot number R (grey are daily values and black are 1-year means) and the homogeneous $aa_H$ geomagnetic activity index averaged over 1-day and 1-year intervals, $<aa_H>_{1dy}$ and $<aa_H>_{1yr}$. The dates of the eclipses shown are marked by vertical mauve lines. Images b, c and d have been processed by Miloslav Druckmüller of Brno University of Technology and are reproduced here with his kind permission: b is the eclipse of 2009 July 22 observed from Enewetak Atoll, Marshall Islands; c is the eclipse of 2013 November 3 observed from Pokwero, Uganda; and c is the eclipse of 2019 July 2 observed from Tres Cruses, Chile. Panels d, e, and f show images of eclipses at the start, maximum and end of cycle 14 that were recorded onto photographic plates and then transcribed onto paper by William Henry Wesley to enable reproduction: f is the eclipse of 1901 May 18 observed from Pamplemousses, Mauritius; g is the eclipse of 1914 August 21 observed from Minsk, Russian Empire; h is the eclipse of 1905 August 30 observed from Sfax, Tunisia. Details of the observers and the equipment used were given in Dyson's 1927 paper. Wesley was able to use photographs taken at different exposure levels to record structure over a range of distances from the Sun, something that is also achieved by modern image processing.*

# The eclipse of 14 December 2020

An excellent example of what can be detected with modern cameras was during the eclipse of 14 December last year, shown in Figure 1. It reveals a coronal mass ejection (CME) leaving the Sun down a classic sunspot-minimum streamer belt. CMEs are large eruptions of the solar corona which, if they hit Earth's magnetosphere, can drive major space weather events. The left hand panel of Figure 4 shows the path of totality in this eclipse as a mauve line, with black dots giving the location of totality at 30 minute intervals. The location marked M is where the image shown in Figure 1 was taken. It can be seen that this event was over land for under 30 minutes.

When he first saw this image, our colleague Luke Barnard, who with Chris Scott runs the highly successful "Solar Stormwatch" citizen science space weather project to track CMEs from the Sun to the Earth (*Barnard*, 2015a; b), asked "what are the chances of that?!" It is a very good question. At sunspot minimum, CMEs visible from Earth using a coronagraph take place at the rate of about one every three days (*Lamy et al.*, 2019) and viewing with polarising light we could detect a CME for about two hours, making our chances in the few minutes viewing afforded by an eclipse about 1 in 36. Given total eclipses occur at a rate of about 2 every 3 years that means we should see an event like this roughly every 24 years, with modern camera technology. At sunspot maximum, the CME rate rises to about 10 per day and this would see suggest that we should see a CME in just about every eclipse at sunspot maximum. However, studies show that CMEs that are seen in spacecraft coronagraph data during total eclipses are not easy to detect in the eclipse images although, with knowledge of the coronagraph image, common features can be identified (e.g., *Pasachoff et al.*, 2015) and in some cases precursors (e.g., *Koutchmy et al.*, 2002) or after-effects (*Pasachoff et al.*, 2009) of CMEs have retrospectively been identified in eclipse images. A major factor here is the lack of temporal information in an eclipse snapshot, that being information that greatly aids the detection of CMEs in coronagraph data. That this is a key factor in why CMEs are so rarely seen in eclipse data was demonstrated by experiments that look at the differences between eclipse images taken at different points along the path of totality: for example *Hanaoka et al.* (2014) were able to detect a CME this way in the difference between images taken 35 minutes apart during the same eclipse, although the event was not readily identified in either image on their own. This is why projects like *Solar Stormwatch* use differenced images to detect CMEs.

Historically, the first observation of a bust of dense material ejected from the Sun was made using 80 MHz radiowaves by the Culgoora radioheliograph on 1 March 1969

(*Riddle*, 1970). This was a moving type IV radio burst, which is emission from a dense plasma cloud. The first optical detection of such an event was on 13 December 1971, made by the first space-based coronagraph on OSO-7 (*Tousey*, 1973). More events were soon detected from Skylab, which was launched in 1973 (*MacQueen et al.*, 1974). The ARTEMIS-II catalogue lists a total of 39188 CMEs over the interval March 1996 to September 2018, covering a full Hale solar magnetic cycle (see *Floyd et al,* 2013 for a description).  This is an average rate of 1742 per year. Using a conservative assumption that an event could be detected for about one hour gives an average observation probability of 0.199 at a time selected at random.  This would imply a repeat period of CME observations in eclipses of 7.5 years.  In fact, our best estimate of the repeat period for such an event is 160 years because we believe that before 14 December 2020, the only CME detected purely from a individual eclipse observation was on 18 July 1860 (*Eddy*, 1974).

# The eclipse of 18 July 1860

The right-hand map in Figure 4 shows the path of totality of this eclipse, the letters pointing to the locations of various observers discussed here, who are listed in Table 1. Sketches of the corona that they made are presented in Figure 5.

This eclipse was first observed soon after sunrise at Steilacoom, near Tacoma on the Puget Sound (south of Seattle) by James Melville Gillis, an astronomer and US naval officer who founded the United States Naval Observatory. His report describes what we would now recognise to be a normal sunspot maximum corona. Half an hour later it was observed by Edward Ashe, the first director of the Quebec observatory, who had sailed with an American expedition to Cape Chidley, on the East shore of Ungava Bay in Labrador, Canada. Ashe did not have very good observing conditions but did note "a white flame shooting up a considerable distance" - an observation that becomes significant when we note it was in the Sun's southwest quadrant.  About 90 minutes later, totality arrived at the southern coast of the Bay of Biscay, in the Basque region of Spain. Many observers were ready here and most of them record curvature of the streamers in the Sun's southwest quadrant – this is clear in the sketches of Lewis, Murray, Oom, and Galton in Figure 5. There are hints of it in the sketches of Winnecke and Weedon and only Weiler draws only the normal straight streamers.  James Maurice Wilson was a teacher of mathematics and science Rugby School and later headmaster at Clifton College and man who did a great deal to further and improve science education in Britain. His sketch from a mountain-top near Pancorbo is an almost ideal drawing of a CME in that quadrant. Ranyard's review quotes several text descriptions of a Turkish scimitar (as in the sketches of Lewis, Murray, Oom, and Galton) or "stags horns" (as in Wilson's sketch). It is the observations of the southwest quadrant made when the event arrived at the Mediterranean coast of Spain that provide the strongest evidence that this event was a CME.

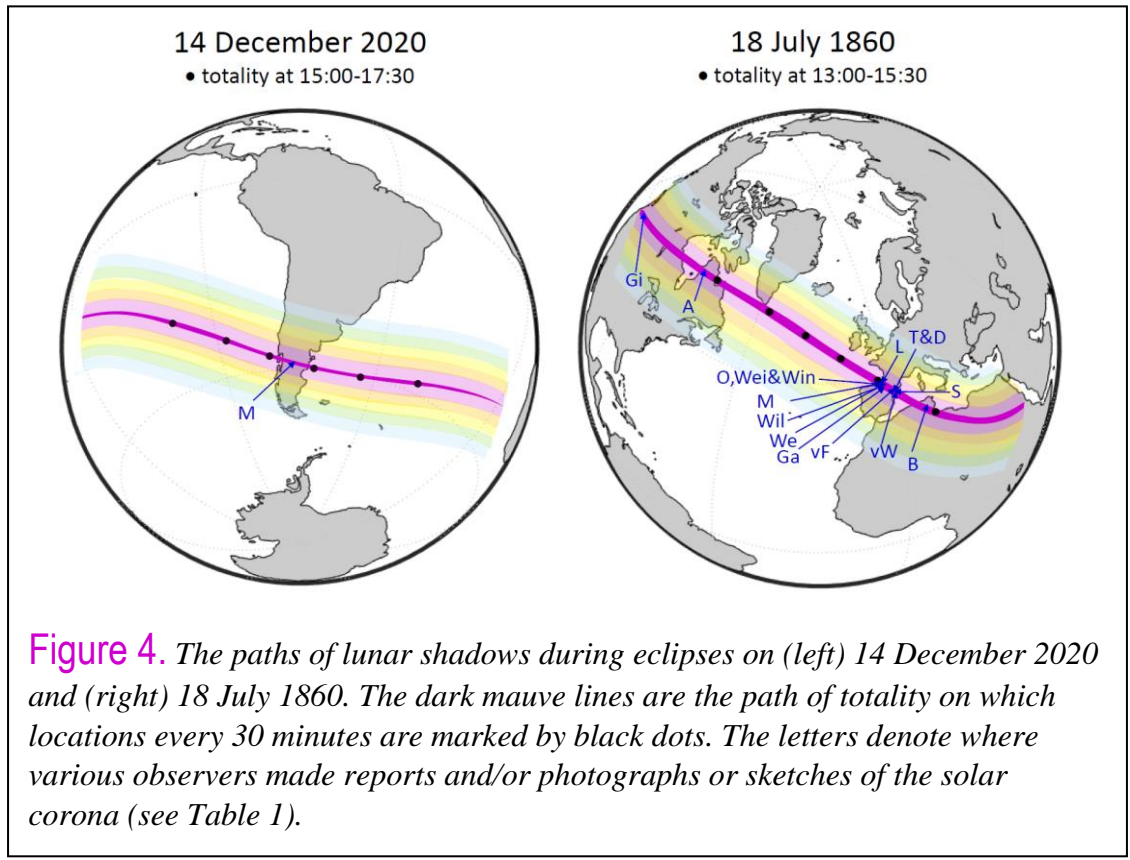

**Figure 4.** *The paths of lunar shadows during eclipses on (left) 14 December 2020 and (right) 18 July 1860. The dark mauve lines are the path of totality on which locations every 30 minutes are marked by black dots. The letters denote where various observers made reports and/or photographs or sketches of the solar corona (see Table 1).*

**Table 1.** *Observers of the 18th July 1860 eclipse discussed in this paper*

| On map | Δt (hr) | Observer | Location | UT of eclipse maximum (h:m:s) | Latitude (degN) | Longitude (deg E) | Duration of totality (m:s) |
|---|---|---|---|---|---|---|---|
| Gi | 0 | Gillis | Steilacoom, near Tacoma, USA | 12:58:31 | 47.17 | -112.59 | 1:48 |
| A | 0.53 | Ashe | East shore, Ungava Bay, Canada | 13:30:19 | 59.61 | -65.42 | 2:59 |
| L | 2.04 | Lewis | Blibao, Spain | 15:01:23 | 43.24 | -2.91 | 2:32 |
| M | 2.05 | Murray | Laudio, Spain | 15:01:30 | 43.14 | -2.96 | 2:48 |
| O | 2.06 | Oom | Pobes, Álava, Spain | 15:02:02 | 42.86 | -2.97 | 3:12 |
| Wei | 2.06 | Weiler | Pobes, Álava, Spain | 15:02:02 | 42.86 | -2.97 | 3:12 |
| Win | 2.06 | Winnecke | Pobes, Álava, Spain | 15:02:02 | 42.86 | -2.97 | 3:12 |
| Wil | 2.06 | Wilson | Pancorbo, Spain | 15:02:19 | 42.62 | -3.12 | 3:23 |
| We | 2.06 | Weedon | Miranda de Ebro, Spain | 15:02:22 | 42.69 | -2.94 | 3:18 |
| Ga | 2.07 | Galton | LaGuardia, Álava, Spain | 15:02:56 | 42.54 | -2.62 | 3:14 |
| Vf | 2.20 | von Feilitzsch | Castellon de la Plana, Spain | 15:10:02 | 39.98 | -0.05 | 3:13 |
| T | 2:19 | Tempel | Torreblanca, Spain | 15:09:46 | 40.22 | 0.22 | 3:16 |
| D | 2:19 | Donati | Torreblanca, Spain | 15:09:46 | 40.22 | 0.22 | 3:16 |
| S | 2.19 | Secchi | Desierto de las Palmas, Spain | 15.09:55 | 40.09 | 0.06 | 3:17 |
| vW | 2.20 | von Wallenberg | Valencia, Spain | 15:10:45 | 39.45 | -0.36 | 1:57 |
| B | 2.41 | Bulard | Lambaesis, Tazoult, Algeria | 15:23:12 | 35.49 | 2.26 | 3:03 |

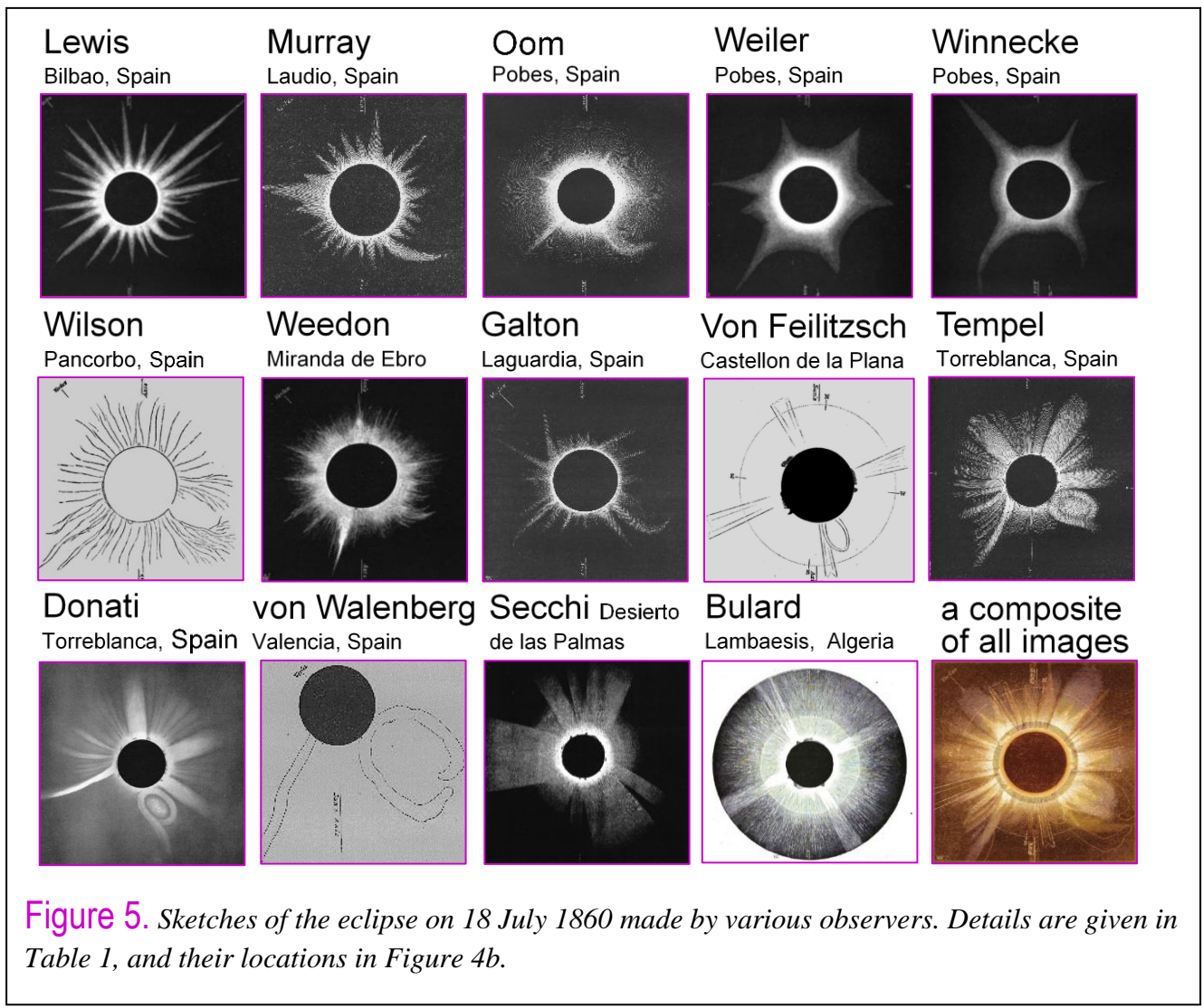

Figure 5. *Sketches of the eclipse on 18 July 1860 made by various observers. Details are given in Table 1, and their locations in Figure 4b.*

Ernst Wilhelm Leberecht Tempel was a prizewinning German astronomer of some note and an accurate and skilful artist. He was a prolific discoverer of comets (discovering or co-discovering a total of 21). A main-belt asteroid and a lunar crater are named after him. In 1860 he had recently arrived at the observatory in Marseille and he joined an Italian scientific expedition to Toreblanca in Spain to observe the eclipse, organized by Giovanni Battista Donati and Francesco Carlini, directors of the Florence and Milan

Observatories, respectively. Donati was the first astronomer to observe the spectrum of a comet and correctly concluded that comet tails contain luminous gas and do not shine merely by reflected sunlight. As mentioned above, he was almost certainly first man to recognise space weather as a distinct discipline. This followed his investigation of the great aurora of 4 February 1872. Sadly, he was unable to pursue the idea as he contracted cholera while attending a scientific conference in Vienna and died in 1873. Given that they were part of the same expedition and published in the same booklet, we cannot regard the images by Tempel and Donati as independent. Nevertheless, both men stated this was the best possible representation of what they saw. The cruder sketches of von Wallenberg and von Feilitzsch are consistent with a the structure in the same quadrant. Bulard observed the eclipse from the Roman archaeological site at Lambaesis in Algeria and although he omitted it from his drawing, his text refers to a structure "like a twisted tulip leaf" in the SW quadrant which is a rather good description of what we often see in the lower corona in the wake of a CME.

The one exception is the drawing showing only radial streamers and is by one of the most famous observers in southern Spain that day, Father Angelo Secchi. Secchi and de la Rue used photography seriously for the first time during this eclipse and were interested in prominences (and proving they were a solar, not a lunar phenomenon) rather than the corona. Secchi's photographs, on which he based his sketch, survive and show that the exposure levels reveal structure very close to the lunar limb but almost nothing about the more distant corona.

# The solar corona during the Maunder minimum

In terms of space weather, the 1860 eclipse is not of any great significance in itself. However, it demonstrates one vitally important fact. That is that good observers, with only the most rudimentary equipment, can observe structure in the solar K corona. It has often been argued that observers detect streamers in the solar corona because that is what they expect to see. It is an argument that ignores the great image processing ability of the human brain that can often detect structures that can be lost in photographs because of exposure levels and the great dynamic range of the intensity of light scattered by the corona. None of the observers in 1860 had any idea what a CME was, and yet they recorded what they saw. The 1860 event shows that many observers, including amateurs with only the most basic equipment, are able to detect and record coronal structure if it is present.

This goes to the heart of one of the great debates about the solar corona – namely what was it like during the Maunder minimum? In his seminal paper on the Maunder minimum, published in 1976, Jack Eddy drew on text descriptions that failed to mention any structure in the solar corona during eclipses that took place in the Maunder minimum.  Some of these were detailed and precise descriptions from trained and skilful observers, such as French mathematician and cartographer Jean de Clapiès and astronomer François de Plantade who observed the eclipse of 12 May 1706 from the Babotte tower in Marseille. Nevertheless the problem with this argument is that "the absence of evidence is not evidence of absence". Incidentally that quote has been used by great astronomers, such as former RAS (and RS) president Martin Rees and Carl Sagan; however, the quote appears to have been first used in 1891 by a William Housman in a letter to the Live Stock Journal of London (about a bull that had reputedly gone missing!).  Many critics of Eddy's evidence use the argument put forward by Mabel Loomis Todd in her influential 1894 book '*Total Eclipses of the Sun*', namely that Maunder minimum observers did not report coronal structure because they could not see it. This underestimates the skill of the best observers in the early 18[th] century: *Hayakawa et al*. (2021) show that some modern naked-eye observers can accurately reproduce coronal structure and there are many ancient texts that refer to what we now call the corona. The problem with these is that they were necessarily expressed in terms of analogies that can be interpreted different ways.  Hence, the best are the simplest – for example an inscription on an oracle bone from the Shang dynasty in China (1751-1111BC) says simply "dawn, fog, three flames ate the Sun" (*Wang and Siscoe*, 1980) which appears to be a wonderful description of a sunspot-minimum eclipse. We conclude that coronal structure could and would have been detected by the best Maunder minimum observers and by farthe most likely interpretation is that its was not recorded because it was not there.

What Eddy did not have were any sketches or paintings of a Maunder minimum eclipse to support his argument.  At the time he was writing, it was known that an astronomer in Nürenberg, working at her father's well-equipped observatory on the bastion of the castle had observed and, in real time, painted the eclipse.  However, the paintings were lost after World War 2 and there was no knowledge of what they portrayed.  The astronomer was Maria Clara Eimmart and she was a well-trained and highly-skilled observer, astronomer and artist.  It is instructive to compare her painting of the moon, made in 1697, with that made by famous astronomer Jean Dominique Cassini in the same year.  Eimmart's is more accurate in many ways. For example, she accurately records the large crater Tycho and its associated ray system (radial streaks of fine ejecta thrown out during the formation of an impact crater) whereas both are missing from Cassini's representation. Furthermore, and of great relevance to her eclipse painting, Eimmart's depiction looks like what one actually sees, whereas Cassini uses stylized representations of craters, mares and mountains. In another painting of the half moon, Eimmart accurately recorded horizon glow on the dark limb of the moon which was only confirmed as a real phenomenon by the Apollo missions. Eimmart often painted onto blue card using yellow paint which she applied more sparingly for fainter features.

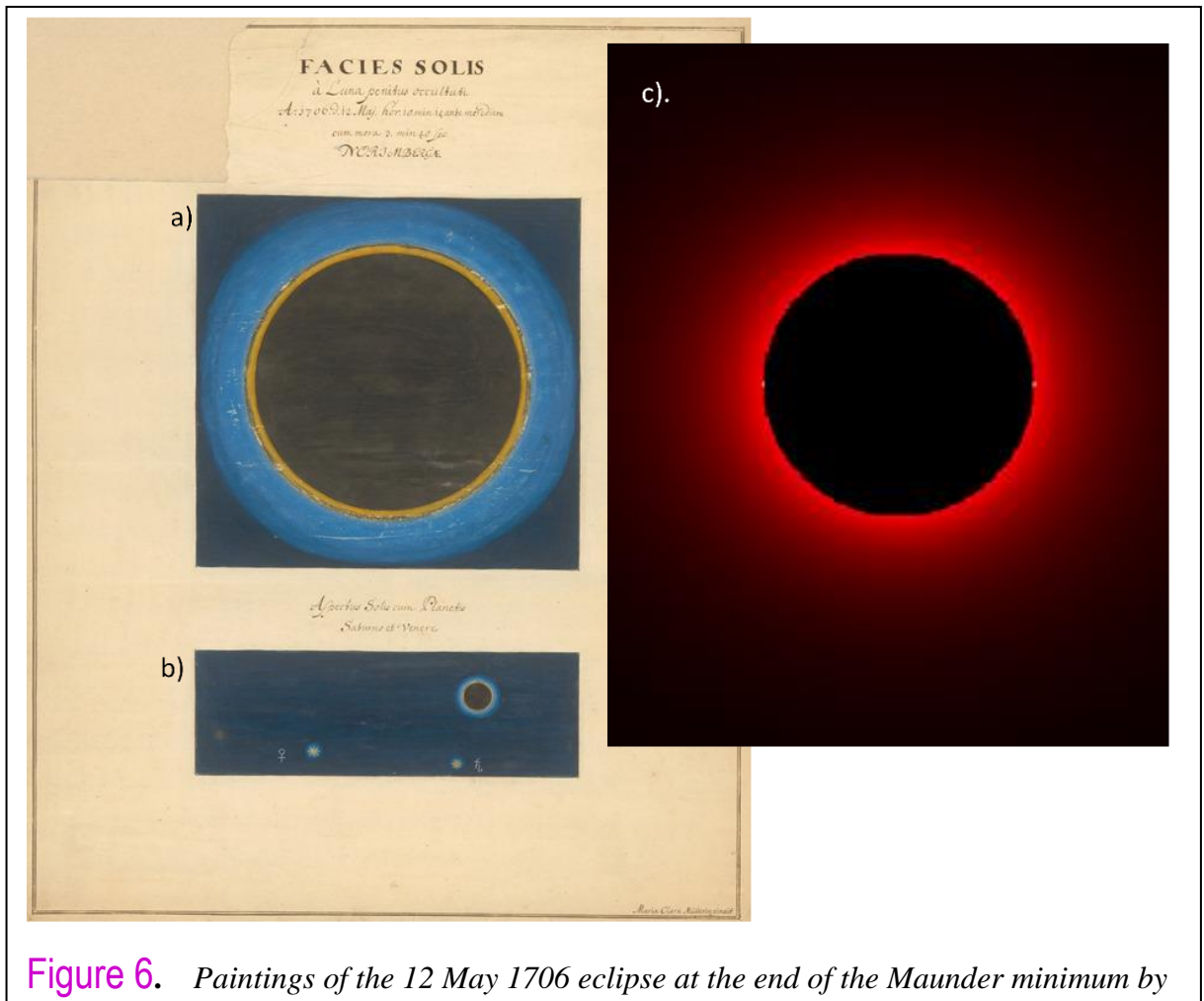

**Figure 6.** *Paintings of the 12 May 1706 eclipse at the end of the Maunder minimum by Maria Clara Eimmart, as observed from Nürenberg observatory. (a) shows the structureless corona (b) places the event in the context of the sky with Saturn, Venus and Mercury accurately positioned. (c) shows the eclipse predictions by Riley et al (2015) from the MAS MHD model of the corona for radically reduced open solar flux during the eclipse, such that the F-corona dominates the K-corona.*

In 2012, Eimmart's eclipse paintings that were thought lost were re-discovered by Markus Heinz of the Staatsbibliothek zu Berlin. They are shown here in Figures 6a and 6b, along with a prediction of the Maunder minimum corona by *Riley et al.* (2015) shown in Figure 6c. The main painting (a) shows a faint structureless corona with a bright ring close to the solar limb while the second paining has a wider view and accurately positions Saturn, Venus and Mercury relative to the moon and Sun. In correspondence, guests who were present at the Nürenberg observatory that day, such as eclipse mapmaker Johann Gabriel Doppelmeyer, praise Eimmart for having captured what they also saw. Another observer in Nürnberg, Johann Philipp von Wurzelbauer, depicted the bright ring but did not report or draw any light from the more distant corona. Eimmart's painting is in perfect accord with clear and precise text descriptions, in particular those by Jean de Clapiès and François de Plantade who observed the event from Marseille. The prediction by *Riley et al.* (2015) uses a small, but non-zero magnetic field threading the photosphere, which results in the F-corona dominating over the K-corona at almost all radial distances from the lunar limb. Further analysis of this, and some other depictions of this event (of a more commercial rather than an astronomical nature), and a comparison with depictions of the 1715 eclipse, are discussed by *Hayakawa et al.* (2021).

The paintings turned out to be Eimmart's last major contribution to astronomy. Like several other women scientists in Germany at the time, she shrewdly chose a husband who would allow her to keep working (*Bernardi*, 2016) and in the year of the eclipse she married her father's student, Johann Müller, who was being trained to be the next director of the observatory and with whom Maria Clara had a good and close working relationship. Tragically, the plan failed as she died in childbirth at the age of just 31, just a few months after the eclipse.

# The application of eclipse observations

These studies have provided a way in which historic eclipses can contribute to our understanding of past space climate change and to constraining our models. From modelling of the open solar flux that uses sunspot numbers to quantify the magnetic flux that emerges through the solar photosphere, *Lockwood and Owe*ns (2014) and *Owens et al.* (2017) modelled the total latitudinal width of the streamer belt on each solar limb, $\Lambda_{SB}$, and of the coronal holes $\Lambda_{CH}$ (where $\Lambda_{SB} + \Lambda_{CH} = \pi$). Some results are shown in Figure 7.

In the collection of events at low sunspot numbers we have added Eimmart's painting of the 1706 eclipse and the cross-like structure in the 1715 event seen from Cambridge and recorded in an anonymous sketch that Roger Cotes sent to Newton (which he describes as "by a very ingenious Gentleman representing the appearance as seen by himself"). In the sunspot maximum images along the bottom we have added Don José Joaquín de Ferrer's eclipse drawing (made at Kinderhook, New York state, USA) of the total eclipse of 16 June 1806, one year after the peak of one of the two weak sunspot cycles during the Dalton minimum (*Hayakawa et al.*, 2020). Also shown are Tempel's depiction of the 1860 event and Lord (James Ludovic) Lindsay's photograph of the solar maximum corona during the eclipse of 12 December 1871, observed from Baiku, India - one of the first photographic prints to capture the full nature of the corona.

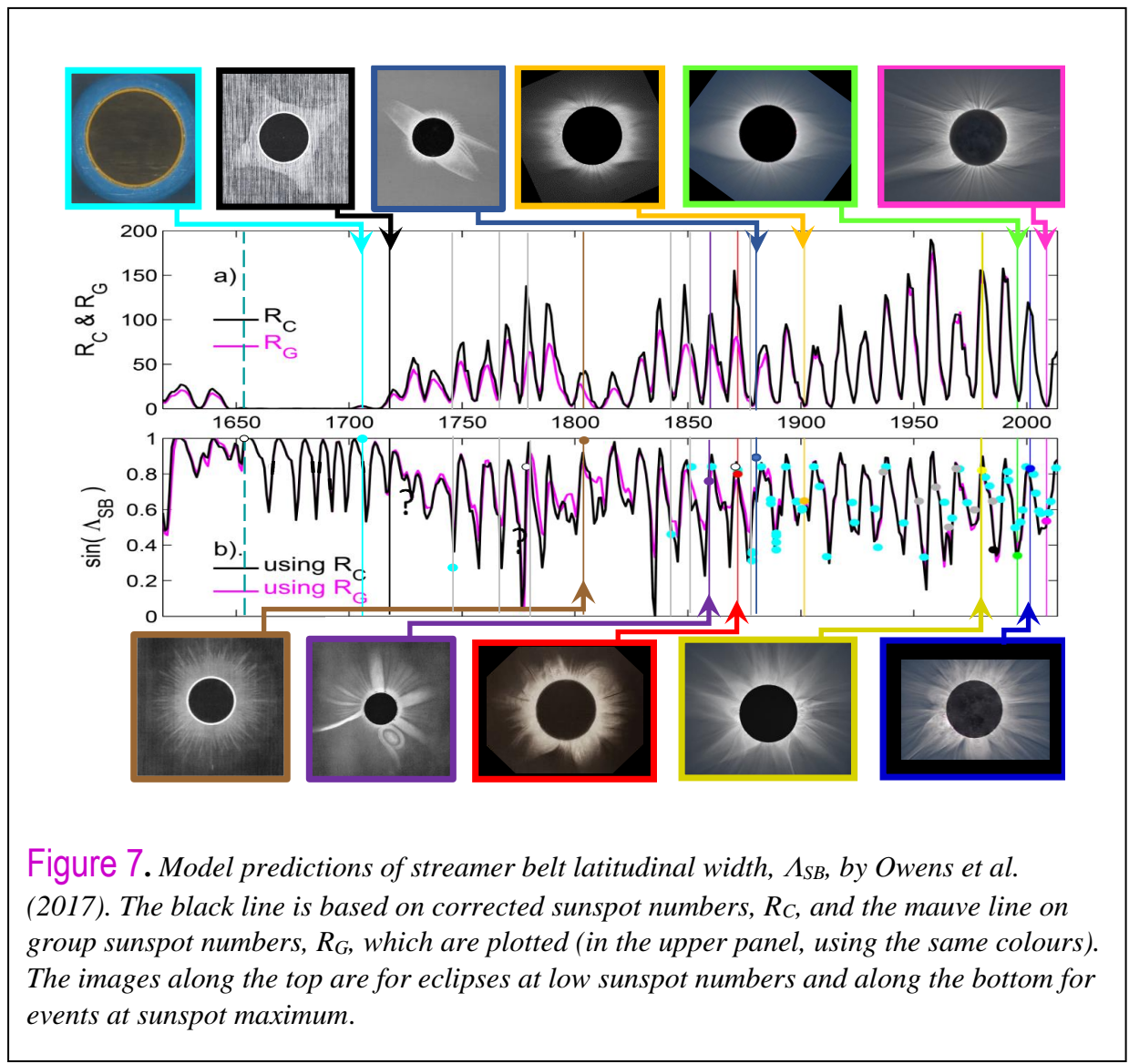

Figure 7. *Model predictions of streamer belt latitudinal width, $\Lambda_{SB}$, by Owens et al. (2017). The black line is based on corrected sunspot numbers, $R_C$, and the mauve line on group sunspot numbers, $R_G$, which are plotted (in the upper panel, using the same colours). The images along the top are for eclipses at low sunspot numbers and along the bottom for events at sunspot maximum.*

The dots give the total latitudinal width of the streamer belt (the average for the two limbs) from the eclipse images. For modern eclipses these estimates agree exceptionally well with both modelled values, values from coronagraph data and values inferred from the divergence of coronal magnetic field lines, modelled from photospheric magnetograph data (*Owens et al.*, 2017).

There are some important points to note about Figure 7. There has been much debate about the calibration of the sunspot number record, but the modelled open solar flux and streamer belt width turn out not to be critically changed by the differences between the various composites of sunspot observations.  In addition, we can discount persistent (and in our view unfounded) suggestions that the Maunder minimum was no deeper than the Dalton minimum (see *Usoskin et al.*, 2015) given that a full solar maximum corona was seen during the Dalton minimum, but only a weak structureless corona at the peak of the weak sunspot cycle at the end of the Maunder minimum. The emergence of the Sun from the Maunder minimum is interesting and it is useful to compare the eclipses of 1706 and 1715. The sketches by Roger Cotes (and by his "ingenious gentleman") show a streamer belt in the 1715 event and a weaker belt at right angles to it. This 'cross-like' structure has quite often been reported and sketched in eclipses since (as in some of the sketches of the 1860 event shown in Figure 5). *Hayakawa et al.* (2021) have recently pointed out that the significance of this is that by 1715 the Sun had regained visible streamers whereas none were detected in just 9 years earlier in 1706.  Both the 1706 and 1715 events were near the peaks of weak sunspot cycles at the end of the Maunder minimum, However, the sunspot cycle around 1715 is very different from that around 1706, and not just in its greater magnitude. In 1706 virtually all the sunspots were in the southern solar hemisphere, as they were throughout the Maunder minimum, whereas by 1715 spots were almost equally shared between the two hemispheres. Furthermore, the cycle around 1706 did not show the famous "butterfly wing" pattern, with spots migrating to lower heliographic latitudes as the cycle progresses - something that has been seen in all the sunspot cycles since the Maunder minimum for which we have heliographic latitude data on sunspots. For the cycle around 1715, the butterfly pattern was observed in sunspot latitudes in the data from the Meudon observatory.

Figure 8 places the two cycles (and eclipses) at the end of the Maunder minimum in context by comparing with data from modern sunspot cycles (numbers 9-24, spanning the years 1840-2019). The plot shows the peak open solar flux (OSF) in each of these cycles, computed, with uncertainties, from geomagnetic activity data (*Lockwood et al.*, 1999, 2014) as a function of the peak group sunspot number, $R_G$, for the same cycle, both being annual means.  There is a clear relationship between the two. Perturbing the OSF values at random (using a Gaussian distribution of standard deviation equal to the estimated OSF uncertainty) and fitting with a $3^{rd}$-order polynomial in $R_G$ was repeated 10,000 times and the grey area gives the 2-sigma uncertainty range of the ensemble of the 10,000 fits at each $R_G$. The black line is the median of that ensemble.

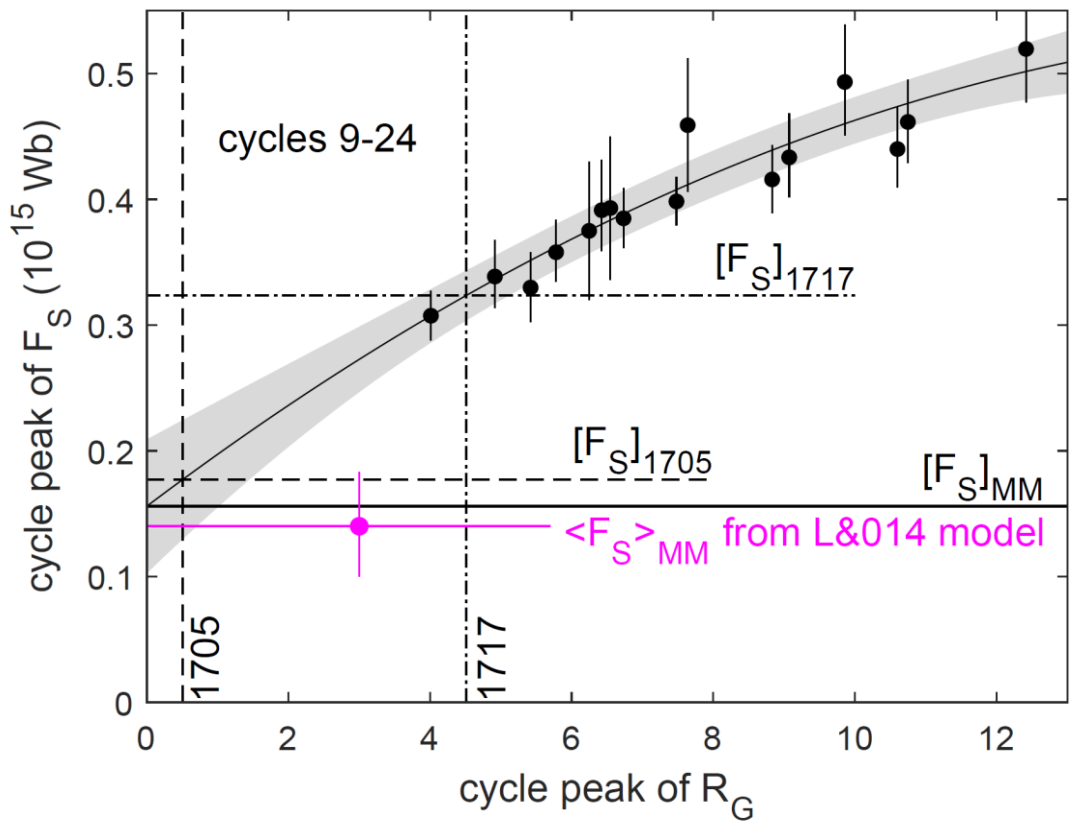

**Figure 8.** *An empirical check on the space climate model made by Hayakawa et al. (2021). The points show the peak open solar flux, $F_S$, derived from 4 combinations of different geomagnetic activity indices by Lockwood et al. (2014) as a function of the peak group sunspot number, $R_G$, compiled by Vaquero et al. (2016) for solar cycles 9-24. The black line is a $2^{nd}$-order polynomial fit to these data and the surrounding grey area is plus and minus the error in this fit, at the 2-sigma level. Extrapolating this simple empirical relationship to $R_G= 0$ gives a value for the OSF in the Maunder Minimum, $[F_S]_{MM}$ which is consistent with the average for the MM from the modelling of Lockwood and Owens (2014) and Owens et al. (2017), shown by the mauve line. (Note the vertical mauve line gives the range of variation in this modelled value caused by the model due to including a continuation of the cyclic variation in the OSF loss rate in the Maunder Minimum). From the empirical fit are scaled the values of the open flux at the peak of the extremely weak and disordered sunspot cycle at the end of the Maunder minimum around 1705, $[F_S]_{1705}$ and at the peak of the much stronger and more ordered cycle that followed it which peaked in 1717, $[F_S]_{1717}$. This is the cycle in which another eclipse occurred in 1715 and is like a modern sunspot cycle in terms of its OSF levels, sunspots numbers, sunspots being both solar hemispheres and showing a butterfly progression of spots from high to low heliographic latitudes. Whereas the 1706 eclipse revealed no streamers, they were seen in 1715 and all subsequent events.*

Extrapolated down $R_G = 0$ these fits give an OSF in the Maunder minimum of $[F_S]_{MM} = 1.5 \times 10^{14}$ Wb which agrees exceptionally well with the average value for the Maunder minimum predicted by the model and is just under one third of the largest value known (which was for solar cycle 19, between 1954 and 1964). Also shown are the values obtained for 1705 and 1717 which are the dates of the maxima in the OSF close to the eclipse dates, as deduced from the dates of the minima in cosmogenic isotope records. It can ben see that for the cycle around the 1706 eclipse, the open flux had hardly risen above the value inferred for the Maunder minimum whereas the value for the next cycle, around the 1715 eclipse, it had returned to within the range seen since 1840 (the lowest value in both peak OSF and peak $R_G$ being set by cycle 12 between 1879 and 1890).

# The way forward

Modern modelling and modern observations continue to improve and enable our reconstructions of past space weather. The results of the Ulysses spacecraft measurements of the heliosphere field out of the ecliptic plane were crucial in bringing historic geomagnetic observations to bear on the problem (*Lockwood et al.*, 1999) and results from Solar Orbiter and Solar Probe promise to bring further advances. The modelling by *Owens et al.* (2017) that covers the last 400 years, uses modern in-situ spacecraft observations and the state-of-the-art "Magnetohydrodynamics Around a Sphere" (MAS) global coronal model that is constrained by photospheric magnetic field observations that have also been made during recent solar cycles. For a given Carrington rotation (CR), the MAS model extrapolates the photospheric field distribution outward to 30 solar radii while self-consistently solving the plasma parameters on a non-uniform grid in polar coordinates, using the MHD equations, with the vector potential $\vec{A}$ used to ensure current continuity is conserved ($\vec{\nabla}.\vec{J} = 0$). Note that the MAS model was also used in the predictions of the appearance of the corona during the Maunder minimum shown in Figure 6c. Improvements in the numerical modelling will, like further observations and enhancements in our understanding, continue to improve the reconstructions. The increased use of physical principles, rather than empirical statistical relationships, means we have more confidence in the reconstructions although, as in Figure 8, the latter can still be used to check the modelling. In addition, the reconstructions also benefitted greatly from the exceptionally long and low sunspot minimum (for the modern era) of 2009, which extended the range of conditions for which we have spacecraft observations of the interplanetary medium. Re-analysis techniques are now allowing us to use model simulations to extrapolate reconstructions between the historic data available. In addition, data assimilation and model ensembles are also showing great promise in exploiting solar, heliospheric and magnetospheric models to forecast events. There is still much to do for a full and general climatology in relating our growing knowledge of variability of annual means to the statistical distribution of events on timescales of a day and shorter. However, Donati's visionary concept of cosmic meteorology is now giving increasing benefits and applications.

## Space Climate
### Beyond Sunspots

We can compare variations in space weather over the last 400 years with those over the past 9000 years using the abundances of cosmogenic radionuclides. These are isotopes that are found in terrestrial reservoirs, such as tree trunks and ice sheets, that were generated in nuclear reactions caused by atoms in Earth's atmosphere being hit by galactic cosmic rays. Earth is partially shielded from cosmic rays by the magnetic field that is dragged out of the Sun by the solar wind, filling the heliosphere. This heliospheric field is enhanced when sunspot activity is high, a connection that we have learned to quantify and predict using models (*Mackay and Lockwood*, 2002), and this causes the cosmogenic isotope production rate to decrease in a way we can exploit quantitatively (*Usoskin*, 2008). By taking a core toward the tree centre or down into the ice sheet, we can look back in time and see how the solar magnetic field has varied.

The records of cosmogenic isotopes, like much of the information available to us, is of one-year resolution at best. For example, from historic observations of geomagnetic activity we can reconstruct solar wind speed, the open solar magnetic flux that leaves the top of the solar atmosphere and the strength of the heliospheric magnetic field of solar origin that reaches Earth (*Lockwood et al.,* 2014) and hence power input into the Earth's magnetosphere (*Lockwood et al.*, 2017). However, we can only do this at one-year resolution because that averages out the variability in unknown parameters – in particular, the orientation of the near-Earth heliospheric field which, through the mechanism of magnetic reconnection, controls energy input into Earth's magnetosphere, and hence geomagnetic activity, on timescales up to several tens of days (*Lockwood et al.*, 2017). We can make annual reconstructions back to the middle if the 19$^{th}$ century when reliable, well-calibrated magnetometer observations began in sufficient locations around the globe. But variations in annual means can only tell us a limited amount about the most damaging space weather phenomena that come in storms that are typically about one day in duration. The one set of data that we have of this time resolution that covers all years between the Maunder minimum and the grand solar maximum are telescopic sunspot observations.

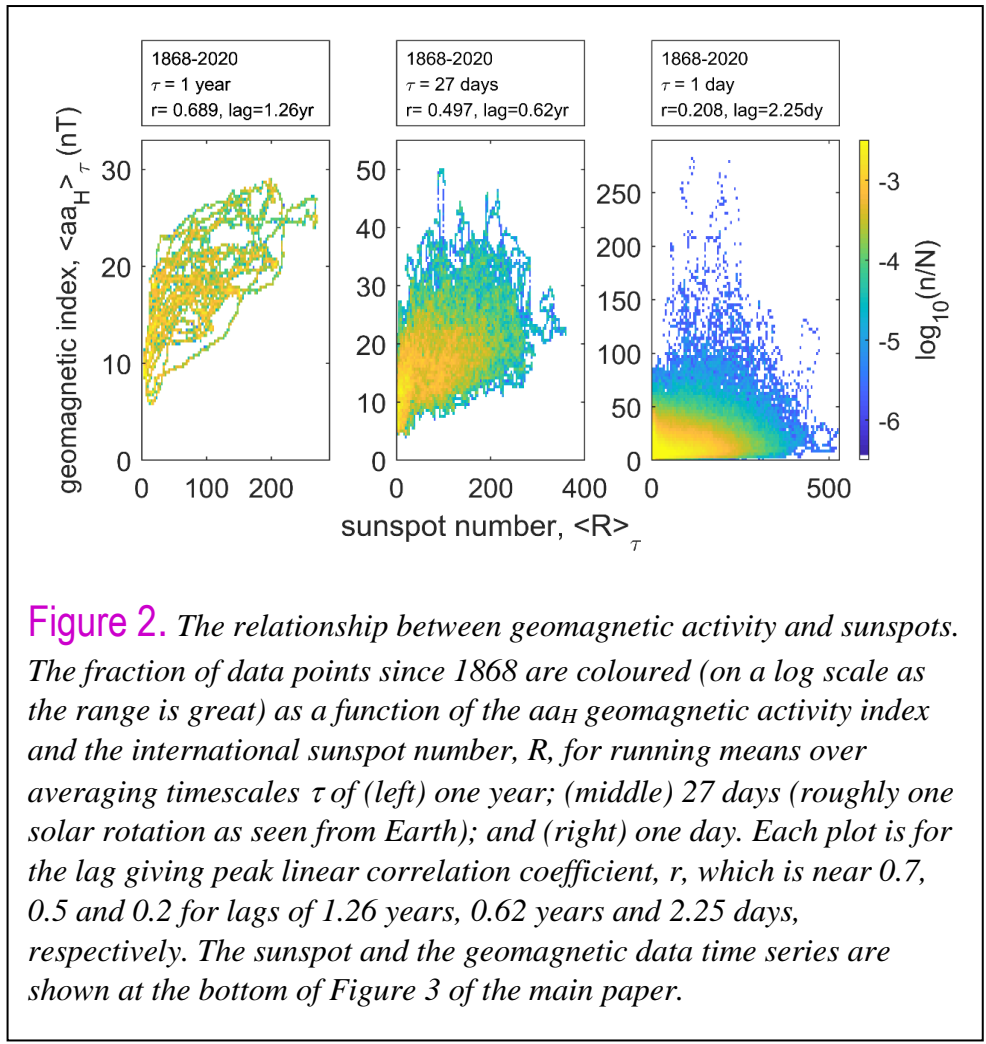

Figure 2. *The relationship between geomagnetic activity and sunspots. The fraction of data points since 1868 are coloured (on a log scale as the range is great) as a function of the $aa_H$ geomagnetic activity index and the international sunspot number, R, for running means over averaging timescales $\tau$ of (left) one year; (middle) 27 days (roughly one solar rotation as seen from Earth); and (right) one day. Each plot is for the lag giving peak linear correlation coefficient, r, which is near 0.7, 0.5 and 0.2 for lags of 1.26 years, 0.62 years and 2.25 days, respectively. The sunspot and the geomagnetic data time series are shown at the bottom of Figure 3 of the main paper.*

The relationship between sunspot cycles and cycles in geomagnetic activity was first noted by Edward Sabine in 1852, but as illustrated by Figure 2, it is only a lose connection. For annual averaging timescales there is a clear correlation, but the scatter is large and so the correlation coefficient is still only 0.7. This is at a lag of 1.26 years and so sunspot numbers do give us some predictive information about geomagnetic disturbance in the near future. However, the correlation falls radically with the averaging timescale used and is only 0.2 for the daily timescale relevant to space weather storms (at a lag of 2.25 days). The right hand panel of Figure 2 shows that although the biggest geomagnetic storms since 1868 have not occurred when sunspot numbers were very low, neither have they occurred when they are very high: rather, they have occurred at low and middling sunspot numbers. Sunspot numbers are a resource that we need to use, being the best and longest record of observations relevant to the near-Earth space environment that we have, but on their own what they tell us is of limited value when it comes to individual storms. Instead, we are learning to apply modern scientific knowledge and models with re-analysis techniques to reconstruct the past behaviour of near-Earth space and to make forecasts for the future (*Barnard et al.*, 2011).

One of two sets of historic observations we have that could potentially tell us about centennial-scale changes in extreme space weather events are observations of aurorae at unusually low latitudes (*Lockwood and Barnard*, 2015). However, these observations are difficult to evaluate as many factors can cause events to remain unrecorded (of which clouds are just one) and changes in the Earth's magnetic field mean that the occurrence patterns of aurora have drifted, relative to centres of population and that has changed the probability of detection. Hence the events of low-latitude aurorae, like the one that set Donati thinking, can contribute, but can only supply one or two pieces of the jigsaw puzzle we are trying to solve. The other historic observations are of the solar corona during total solar eclipses and, as explained in the main text, in 1860 Donati was one of several astronomers who made an intriguing observation of an eclipse of relevance to space weather.

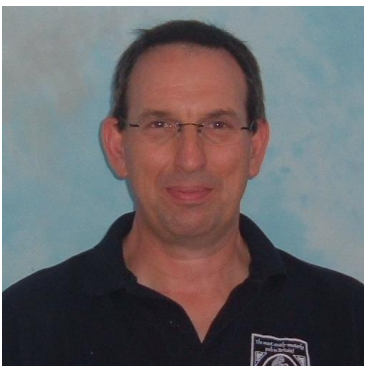

### Prof. Michael Lockwood

is professor of space environment physics, University of Reading, UK. His favourite astronomical object is Earth's magnetosphere and he is grateful to: ESA and NASA for many spacecraft, guitars for being a challenge, marmalade for being nice to eat, his family for everything and science for opportunity

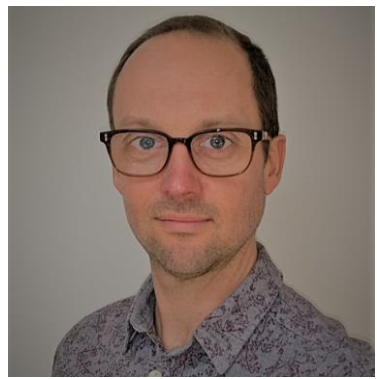

### Prof. Mat Owens

is a space physicist impersonating a meteorologist at the University of Reading. He likes the Sun, running and watching cartoons with his kids